\documentclass{ecai}
\usepackage{times}
\usepackage{graphicx}
\usepackage{latexsym}
\usepackage[caption=false]{subfig}

\begin{document}

\title{Opening the Software Engineering Toolbox for the Assessment of Trustworthy AI}

\author{
Mohit Kumar Ahuja\institute{Simula Research Laboratory, Dept. of Validation Intelligence for Autonomous Software Systems, Oslo, Norway, \{mohit, bachir, pierbernabe, mathieu, arnaud, chhagan, dusica, sagar, aizaz, helge\}@simula.no Funding: This work has received funding from the European Union under grant agreement no. 825619 (AI4EU), the EU landmark project to develop a European AI on-demand platform and ecosystem. Copyright \textcopyright 2020 for this paper by its authors. Use permitted under Creative Commons License Attribution 4.0 International (CC BY 4.0).} \and
Mohamed-Bachir Belaid\footnotemark[1] \and
Pierre Bernab\'{e}\footnotemark[1] \and\\
Mathieu Collet\footnotemark[1] \and
Arnaud Gotlieb\footnotemark[1] \and
Chhagan Lal\footnotemark[1] \and\\
Dusica Marijan\footnotemark[1] \and
Sagar Sen\footnotemark[1] \and
Aizaz Sharif\footnotemark[1] \and
Helge Spieker\footnotemark[1]
}

\maketitle

\begin{abstract}
  Trustworthiness is a central requirement for the acceptance and success of human-centered artificial intelligence (AI).
  To deem an AI system as trustworthy, it is crucial to assess its behaviour and characteristics against a gold standard of Trustworthy AI, consisting of guidelines, requirements, or only expectations.
  While AI systems are highly complex, their implementations are still based on software.
  The software engineering community has a long-established toolbox for the assessment of software systems, especially in the context of software testing.
  In this paper, we argue for the application of software engineering and testing practices for the assessment of trustworthy AI.
  We make the connection between the seven key requirements as defined by the European Commission's AI high-level expert group and established procedures from software engineering and raise questions for future work.
\end{abstract}

\section{INTRODUCTION}

Artificial Intelligence (AI) has increasing relevance for many aspects of the current and future everyday life.
Many of these aspects interfere directly with the personal space of humans, their perception, actions, and, more generally, their data, both online and offline.
Due to this close integration, it is therefore crucial to develop the AI systems in a human-centered fashion such that they are trustworthy and can be accepted by providers, who develop and deploy the AI systems, users, who operate the AI systems, regulatory bodies, who oversee the usage and effects of the AI systems, and affected humans, who are act in cooperation with or next to the AI systems or who's data is subject to processing via the AI systems.

% What belongs to trustworthy AI
To define the extent and more specific definition of a trustworthy AI, a high-level expert group (AI-HLEG) that was set up by the European Commission, identified guidelines and requirements for an AI system that need to be sufficiently fulfilled to be regarded as trustworthy~\cite{AIHLEG2019-EthicsGuidelines}.
On the highest level, an AI system is deemed trustworthy if it behaves according to four ethical principles: respect for human autonomy, prevention of harm, fairness, and explicability~\cite[p. 12]{AIHLEG2019-EthicsGuidelines}; on a more technical level, requirements have been formulated that are supposed "to be continuously evaluated and addressed throughout the AI system's life cycle"~\cite[p. 15]{AIHLEG2019-EthicsGuidelines}.

% What is the challenge with trustworthy AI
Having a definition of trustworthy formulates a goal for the development of AI systems.
The second step is to evaluate if a system fulfills the definition sufficiently and can be deemed trustworthy.
This evaluation should be transparent and accessible to understand its results, robust and reproducible, and both automated and generic as much as possible to allow a low barrier for application to new AI systems.
Since there is no single trustworthiness criterion or even metric, a single evaluation technique is not sufficient or maybe even possible.
The trustworthiness assessment has to consist of multiple techniques, each appropriate for some of the requirements of trustworthy AI and each robust and mature enough to be reliable.

% What do we propose to the extend of these challenges
Tools and techniques for the assessment of trustworthy AI can be taken from the established methods of software engineering research and especially the subarea of software testing.
For 50 years, these communities have proposed methods for the realization and assessment of large-scale, complex software systems.
While the criteria for trustworthy AI do cover more than technical aspects, the AI system itself is still mostly a software system.
Even though their are differences in their engineering, many of the software engineering principles apply to them or are transferable \cite{Amershi2019,Breck2017}.
Recently, motivated through the recent breakthroughs of AI and especially deep learning, the software engineering community has increased the attention on machine learning, both as a tool within software engineering and an area for the application of software engineering principles.

Through the remainder of this short paper, we argue to open the software engineering toolbox with its wide range of methods for the realization and assessment of trustworthy AI systems.
Following the structure of the key requirements for trustworthy AI \cite{AIHLEG2019-EthicsGuidelines}, we link existing techniques with the goals for the fulfillment of these requirements.
It is important to note, that even though there are already many methods available, the research on trustworthy AI is by far not complete.
Our current toolbox, however, provides a strong starting position but needs adjustments and further experiences to be adapted for the specific characteristics of modern AI.

\vspace{-0.6em}
\section{TRUSTWORTHY AI}
%\vspace{-0.6em}
This section discusses an overview of approaches related to the key requirements for Trustworthy AI from the HLEG's Ethics Guidelines from software engineering and adjacent subfields.
We aim to analyse how to map system engineering onto the requirements, and to show examples for techniques, case studies, that have already been explored.
At the same time allows a discussion of existing techniques to identify where future research is required or areas where the software engineering toolbox might be insufficient to properly address aspects of a given requirement.

% \begin{enumerate}
%   \item Human agency and oversight
%   \item Technical Robustness and safety
%   \item Privacy and data governance
%   \item Transparency
%   \item Diversity, non-discrimination and fairness
%   \item Societal and environmental well-being
%   \item Accountability
% \end{enumerate}

\subsection{Human agency and oversight}
The first of the requirements is the necessity for the AI to support human autonomy and the option for the human to inspect and influence the AI's actions.
Human agency directly affects the collaboration between AI and human and to support this interaction, it is important to take appropriate design measures, such as ergonomic and accessible user interfaces (UI) and an excellent user experience (UX).
Human oversight requires the inspection of the AI decision making, either by having interpretable models or having access to design decision documents, source code, or data, depending on the level of expertise of the inspecting party.

It is also one of the main challenges in AI to find a perfect balance between enhancing human agency and preserving a degree of responsibility \cite{floridi2018ai4people}. 
Some "black box" AI techniques prevent the human from understanding the embraced process and thus prevent him from the control. 
We believe that software engineering and testing frameworks can contribute in achieving a better degree of human understanding and control of AI techniques.
Software testing techniques are often based around the goal to design the simplest test cases to determine a system's quality.
Having these tests for AI systems will improve the ability to understand the AI behaviour and its deviations from it.
While there is already work on applying and adopting current testing techniques on AI \cite{Zhang2020}, future work is required to ease their capabilities and expressiveness for human oversight.

\subsection{Technical Robustness and safety}

\begin{figure}[t]
  \centering
  \subfloat[Differential Testing]{\includegraphics[height=4.25cm]{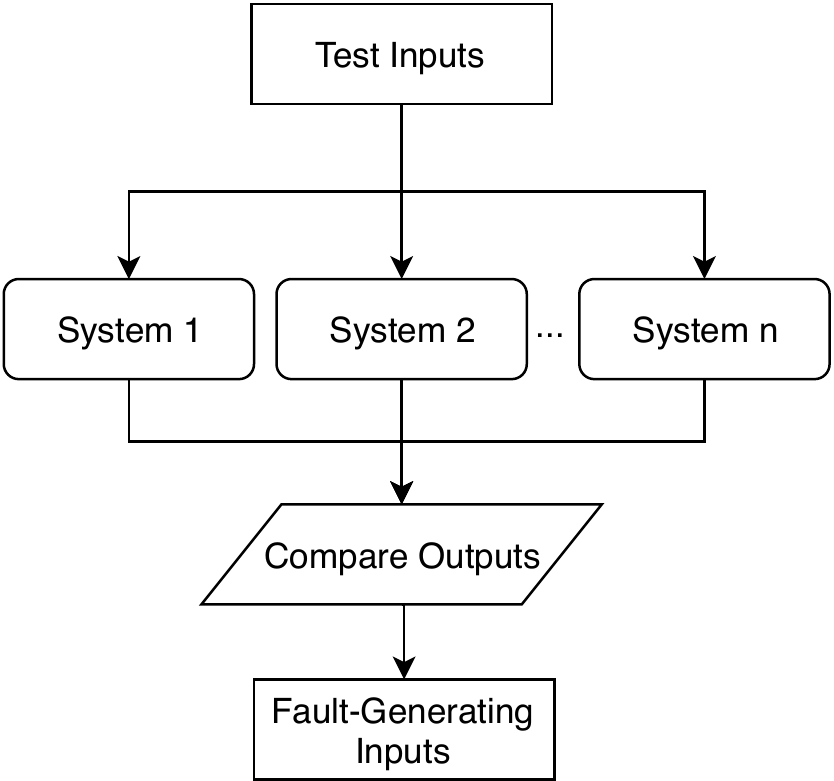}}\hfill
  \subfloat[Metamorphic Testing]{\includegraphics[height=4.25cm]{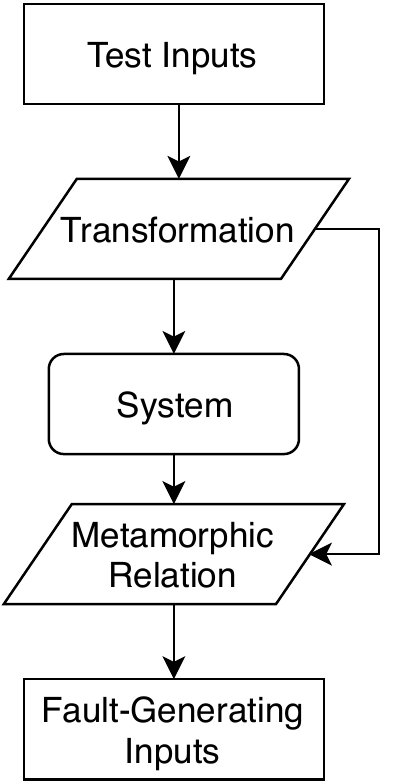}}
  \caption{Schematic Overview of two testing techniques for Deep Learning Systems}
  \label{fig:dl_testing_techniques}
\end{figure}

The technical robustness of AI systems is central to their reliability.
While performing well in their main performance metrics, e.g. the classification accuracy, additional safety, and robustness metrics and the resilience to attacks often remain open challenges \cite{Papernot2018a}.
Of particular relevance are adversarial inputs which are specially crafted to attack an AI system, for example, to cause misclassification or to extract internal information about training data.

These challenges have recently been identified by several software testing techniques and have been adopted towards the testing of AI systems, especially for deep learning.
To highlight two techniques that have been successfully applied towards the testing of deep learning, we briefly discuss \emph{differential} and \emph{metamorphic} testing (see Figure~\ref{fig:dl_testing_techniques}).
In differential testing, a system is evaluated by comparing its behaviour against a set of reference implementations for the same task.
For the same inputs, it is expected that all systems provide similar outputs and if a system diverges it is an indicator of faulty behaviour.
The advantage of differential testing is that the specific test oracles for the inputs, i.e. the precise expected outputs, are not required which allows easier setup of the test cases, especially when defining the test oracles is too costly or complex.
DeepXplore \cite{pei2017deepxplore} first explored differential testing for deep learning.
The paper proposes a controlled way to generate test inputs, similar to adversarial examples, that are likely to identify diverging behaviours and showed promising results on multiple datasets and models.

Metamorphic testing also alleviates the problem of defining precise test oracles.
Here, new test cases are generated with the help of metamorphic relations.
These relations allow to describe a property of the behaviour, e.g. the output, when a change in the input is made.
For example, for an AI-based HR system to rank resumes of applicants, adding relevant keywords should improve the ranking, even though there is no precise definition of the final expected ranking.
In the context of testing AI, metamorphic testing has been applied for testing of autonomous driving systems \cite{Zhang2018}, image classifiers \cite{dwarakanath2018identifying}, or ranking algorithms \cite{Murphy2008}.

\subsection{Privacy and data governance}
% We need to consider two things; 
% a) how can we test that privacy and data governance is preserved
% b) how can we design/test systems without interfering with private user data, e.g. federated learning; testing of systems that "learn at the user's premises" -> here might be inspirations from database testing

% Engineering Privacy / https://ieeexplore.ieee.org/abstract/document/4657365
Privacy protection of individuals who contribute with their personal data towards development of AI is of paramount importance in human-centered AI. 
Any party that curates datasets needs to ensure that the data does not provide means of re-identifying individuals while, at the same time, being effective at predicting patterns of business/societal value. 
Secure data-intensive systems storing personal data typically contain identifying, quasi-identifying, non-identifying and sensitive attributes about individuals. 

Software tools such as ARX \cite{prasser2015putting} can anonymize and perform re-identification risk analysis on large datasets to quantify the risk of prosecutor, journalist, and marketer attacks before the data is used in AI. 
ARX can be used to anonymize data based on criteria \cite{dangi2012privacy} such as k-anonymization (personal attributes are suppressed or generalized until each row is identical with at least k-1 other rows), l-diversity (entails reducing granularity of data), and t-closeness (a refined reduction of granularity by maintaining an underlying data distribution). 
However, in specific cases, quasi-identifying attributes such as the birth date of an individual are required to train AI models. 
Therefore, controlled fuzzification of quasi-identifying attributes \cite{ursin2017protecting} can minimize the risk of re-identification while maintaining underlying patterns of interest in the data. 
For instance, in cervical cancer screening, attributes such as birth date or screening exam date can be perturbed within certain bounds. 
This is primarily due to the fact that the human papillomavirus has an average latency period of 3 months. Therefore, database commands can fuzzy all dates to the 15th of a month (middle), and move months by $\pm 2$ months without affecting disease progression patterns and increasing risk of re-identification.

\subsection{Transparency}
The transparency of an AI system is closely related to its interpretability and explainability, but also to the documentation of its purpose and how it has been designed.
An approach for transparency documentation is the concept of \emph{model cards} \cite{Mitchell2019}, which aims to provide accessible overview of a model for people of different expertise, including all of developers, testers, and technical end-users, similar to a package insert in pharmaceutical products.

Lower level measures for transparency can be achieved via strict traceability and static analysis \cite{urban2019static} to allow the documentation of system behaviour, e.g. in autonomous vehicles \cite{borg2017traceability} in combination with requirements engineering \cite{8491154}.
These techniques allow higher transparency of the AI during development, evaluation, and certification tasks, where they serve mostly technical needs for the development and integration of the AI component.
%Static analysis of data science software \cite{urban2019static}
%Traceability of deep learning systems \cite{borg2017traceability} \cite{bailer2018traceability}

\subsection{Diversity, non-discrimination and fairness}

Adequate diversity in data to train AI systems is necessary to avoid discrimination and maintain fairness in human-centered AI. 
History has taught us that bias in using personal data has harmed several generations of ethnic minorities. 
Lundy Braun \cite{braun2014breathing} reports the implications of biased data in spirometers that measure a person's lung function after a forced exhale. 
The predicted values of a lung's forced vital capacity (in litres of air exhaled) for black people for over a century been lower than white people. 
One of the reasons was that the data was collected from black men working in cotton fields where lint from cotton severely damaged lung function. 
This has resulted in black people receiving very little help from medical insurance companies for several generations.
Even today, race and not socio-economic factor is used as a parameter to predict lung capacity in spirometers used worldwide. 
This unfortunate trend continues in AI systems where a recent study \cite{ledford2019millions} shows that millions of black people are victims of biased decision making in health care systems. 

Data needs to be carefully curated for training AI systems such that variation in human attributes such as different ethnic groups, genders, ages, weights, heights, geographical areas, and medical histories are taking into account for unbiased decision making. 
However, discovering if a data set satisfies all possible combinations of attributes is often computationally intractable. 
Combinatorial interaction testing (CIT) of software has been very effective in finding over 95\% of all faults in a wide range of software systems using a very small set of tests covering all 2-wise/pairwise combinations of features \cite{kuhn2002investigation}. 
CIT has been extended to verify if data in a large relational database contains all pairwise interactions between attribute values of interest \cite{sen2016modeling}. 
Verifying the presence of all pairwise interactions in human attribute values in data set can clarify limitations or guarantee adequate diversity in human-centered AI systems. 

The importance of fairness in software received attention as a dedicated topic in software engineering research \cite{8452913} with close connections for the assessment via software testing methodology \cite{10.1145/3236024.3264838}.

\subsection{Societal and environmental well-being}
%AI systems that interact with people can be seen as socio-technical systems when the social behaviour is measurable. 
%Epidemiology, tracking the use of an AI intervention
%pathways taken by people to keep track. \cite{sen2017portinari}
%self-care revolution IoT and well-being.

Human-centered AI systems need to benefit society and not cause harm. 
It is necessary to see an AI system as not merely a software system but as a socio-technical system where the interaction between people the system is used to evaluate its benefit. Learning from epidemiology, we can evaluate an AI system as if it were an intervention on the public. 
For instance, in \cite{sen2017portinari}, the authors visualize the paths a patient takes after different screening exams for cervical cancer. 
Similarly, there is a need to understand how decisions made by the AI system affect the decisions made by people and the paths they take in life. 
Are people making healthier life choices, environmentally conscious, or giving a helping hand in society after an AI intervention? 
Evidence-based software engineering \cite{kitchenham2004evidence} inspired by epidemiology and clinical studies presents numerous approaches to evaluate the impact of AI on people. 
These approaches include randomized controlled trials, observational studies, and focus group discussions to name a few. 
All these approaches however require careful data collection after a target audience has been exposed to an AI system.

\subsection{Accountability}
Access to personal data used in AI systems should be controlled by its owner in human-centered AI. 
The owner can give consent of use and take away access to personal data whenever he/she wants to. 
This implies that the AI system would need to be re-trained with or without a specific person's data. 
The proof of this operation should be made known to the owner to ensure accountability.
The \emph{blockchain} has the potential to facilitate the accountability of such transactions between a data owner and an AI system. 
The blockchain is a \emph{distributed ledger} which was initially designed to record financial transactions. 
Numerous models of using the blockchain and smart contracts have now been proposed for data access control \cite{drosatos2019blockchain} and AI~\cite{salah2019blockchain}. 
Tal Rapke \cite{linn2016blockchain} suggests that people own and access their health and life record on a decentralized blockchain that does not rely on a central storage facility.
This will liberate organizations from the liability of storing personal data. 
The data will reside on the latest secure technology and using verifiable cryptography and owners of the data will be empowered to decide who they share their data with. 

\section{THE SOFTWARE ENGINEERING TOOLBOX}

\begin{figure}[t]
    \centering
    \includegraphics[width=0.9\columnwidth]{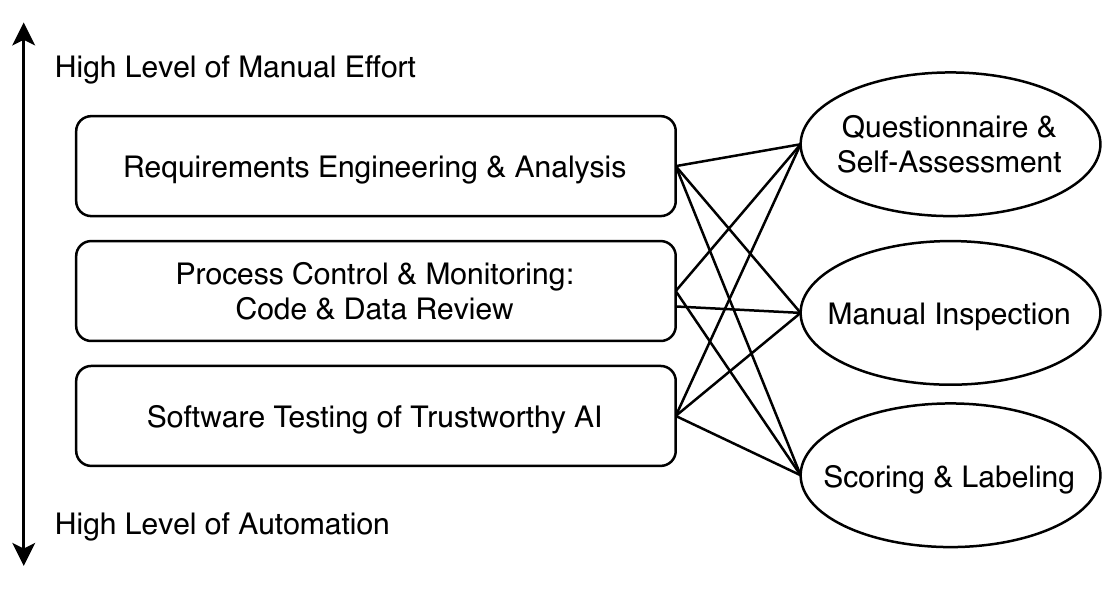}
    \caption{Concepts from the Software Engineering toolbox for the preparation, monitoring and evaluation of trustworthiness in AI projects.}
    \label{fig:setoolbox}
\end{figure}

The discussion of the key requirements on trustworthy AI \cite{AIHLEG2019-EthicsGuidelines} shows that there are many challenges to be addressed, but also a set of methods available that can embraced and extended.
As a general approach towards these challenges, we propose to adopt three main considerations (see Figure~\ref{fig:setoolbox}):
First, since the expectations on trustworthy AI cannot be presented as a strict set of guidelines and rules only, it is recommended to understand their impact on the AI that is developed.
Performing thorough requirements analysis helps to gather these requirements in a systematic way \cite{Belani2019} and to formalize the requirements' impact on the AI including final acceptance criteria and whether they can be assessed automatically or require manual intervention.
One method to guide the requirements analysis at this point could be to formulate checklists for each of the requirements, e.g. similar to this proposition for fairness \cite{Madaio2020}.

%b) process control and monitoring via a combination of manual and automated steps, e.g. CI/CD, code and data review, retrospectives
Second, the realization of a trustworthy AI should be continuously accompanied by regular monitoring instruments.
The goal of this monitoring is to ensure the awareness of trustworthiness measures during development.
These monitoring instruments can include dedicated questions to consider during code and data reviews, as well as retrospective meetings.

%c) automated testing where possible
Third, automated testing should be used to allow automated, repeated, and comparable assessment of trustworthiness.
Where possible, testable acceptance criteria should be defined or test process that can quantify the behaviour of the AI system.
For example, the technical robustness of an AI systems against adversarial inputs can be assessed through automatic techniques.

Finally, in all cases does the qualitative and quantitative summary of the results, e.g. via a score or a badge to attest the quality of an AI system, provides valuable information to the different stakeholder groups, e.g. the providers, their customers, or the users.
A common scoring scheme, similar to the maturity levels in engineering projects, could allow for comparability and accessibility of the results.

\section{CONCLUSION}

The realization of trustworthy AI systems is one of the big challenges for the success of ethical and human-centered AI.
This has been acknowledged by both politics \cite{AIHLEG2019-EthicsGuidelines} and academia.
For the implementation of trustworthiness principles, we argue for the adoption of methods and technologies from software engineering.
Software engineering has a long-standing tradition on the principled construction of complex systems and has already much of the fundamental work available, as shown throughout this paper.

Still, further work is necessary to cover all the requirements on trustworthy AI and to provide the tools and guidelines necessary for the widespread realization of trustworthy AI.
Are the current software engineering tools sufficient to assess AI systems? 
Or do we need to develop dedicated tools?
How can we converge on a set of acceptance criteria for trustworthiness?
How many of the requirements can effectively be assessed in a mostly automated way?
What are the challenges for assessing trustworthy AI by non-specialists or external users? 
How can we present the results in an accessible way?

The software engineering community has already taken up the challenge of software engineering for AI/ML, but often with a focus on the general system engineering, maintenance requirements, and general validation.
However, as the requirements discussed in this paper showed, the engineering efforts need to cast a wider need and address more concerns in the context of trustworthy AI.
This will be an interdisciplinary challenge and the software engineering toolbox can be of central relevance during its development.

\bibliographystyle{ecai}
\bibliography{refs.bib}
\end{document}